\documentclass[twocolumn,twoside,slac_two]{revtex4}
\usepackage{graphicx}
\usepackage{fancyhdr}
\usepackage{xspace}
\usepackage{amssymb}
\usepackage{amsmath}

\def\p0{$\pi^{\rm 0}$}
\def\Angst{$\buildrel _{\circ} \over {\mathrm{A}}$}
\def\de{$^{\circ}$\xspace}
\def\e{\epsilon}

\def\like{\mathcal{L}}

\newcommand{\Fermi}{\emph{Fermi}\xspace}
\newcommand{\fermi}{\emph{Fermi}\xspace}

\newcommand{\hsi}{\emph{RHESSI}\xspace}

\newcommand{\soho}{\emph{SOHO}\xspace}
\newcommand{\stereo}{\emph{STEREO}\xspace}
\newcommand{\sdo}{\emph{SDO}\xspace}

\def\like{\mathcal{L}}

\newcommand{\ltsima} {$\; \buildrel < \over \sim \;$}
\newcommand{\gtsima} {$\; \buildrel > \over \sim \;$}
\newcommand{\lta} {\lower.5ex\hbox{\ltsima}}
\newcommand{\gta} {\lower.5ex\hbox{\gtsima}}
\def\de{$^{\circ}$\xspace}

\def\e{\epsilon}

\begin{document}

\title{Shedding new light on the Sun with the \Fermi LAT}

%

\author{N. Omodei, V. Petrosian, W. Liu, F. Rubio da Costa, Q. Chen}
\affiliation{Stanford University/KIPAC, Stanford, CA, USA 94305}
\author{M. Pesce-Rollins}
\affiliation{INFN-Pisa, Pisa, Italy 56100}
\author{E. Grove} 
\affiliation{U.S. Naval Research Lab, SW Washington, DC, USA 20375}
\author{F. Longo} 
\affiliation{INFN-Trieste, Trieste, Italy 34127}
\author{on behalf of the \fermi-LAT collaboration}

\begin{abstract}
During its first six years of operation, the \Fermi Large Area Telescope (LAT) has detected $>$30 MeV gamma-ray emission from more than 40 solar flares, nearly a factor of 10 more than those detected by EGRET. 
These include detections of impulsive and sustained emissions, extending up to $\sim$20 hours in the case of the 2012 March 7 X-class flares. We will present an overview of solar flare detections with LAT, highlighting recent results and surprising features, including the detection of $>$100 MeV emission associated with flares located behind the limb.  Such flares may shed new light on the relationship between the sites of particle acceleration and gamma-ray emission.
\end{abstract}

\maketitle

\thispagestyle{fancy}


\section{Introduction}
Understanding the processes of particle acceleration and impulsive energy release which occur in numerous sites throughout the Universe is one of the major goals of space physics and astrophysics. 
The Sun is the most powerful particle accelerator in the solar system and its proximity permits investigating the entire electromagnetic spectrum of these acceleration phenomena. 
During solar flares, the Sun is capable of accelerating electrons and ions to relativistic energies on time scales as short as a few seconds, as indicated by observations of X-rays, microwaves, $\gamma$-rays, and neutrons produced when the flare-accelerated particles interact with the solar atmosphere \citep{forr83,kane86}. 
In general, the $\gamma$-ray emission light curve is similar to that of the HXRs (possibly with some delay), lasting for 10--100 seconds. This is referred to as the ``impulsive" phase of the flare. However, the {\it Energetic Gamma Ray Experiment Telescope} (EGRET) on-board CGRO \citep{Kanbach:88,Esposito:99} also detected a sustained emission in gamma rays for more than an hour after the impulsive phases of 3 flares \citep{2000SSRv...93..581R}.
The expected increase of solar activity during the current solar maximum is producing a large number of solar flares, including bright GOES X-class and moderate M-class flares. 

\section{Fermi observations of the Sun}
Launched in 2008, the \Fermi observatory is comprised of two instruments; the Large Area Telescope (LAT) designed to detect gamma-rays from 20~MeV up to more than 300 MeV~\citep{LATPaper} and the Gamma-ray Burst Monitor (GBM) which is sensitive from $\sim$~8 keV up to 40~MeV \citep{GBMinstrument}. During the first 18 months of operation coinciding with the solar cycle minimum, the \Fermi LAT detected $>$100 MeV gamma-ray emission from the quiescent Sun \citep{abdo11}. As the solar cycle approaches it maximum, the LAT has detected several solar flares above 30 MeV during both the impulsive and the temporally extended phases \citep{2011ATel.3635....1O,2011ATel.3552....1O,2012ATel.3886....1T,2012AAS...22042404P,2012AAS...22042403O}.
The first \Fermi GBM and LAT detection of the impulsive GOES M2.0 
flare of 2010 June 12 is presented in \citet{2012ApJ...745..144A}.
The analysis of this flare was performed using the LAT Low-Energy 
(LLE) technique \citep{pela10} because the soft X-rays emitted during the prompt 
emission of a flare penetrate the anti-coincidence detector (ACD) of the LAT 
causing a pile-up effect which can result in a significant decrease in 
gamma-ray detection efficiency in the standard on-ground photon analysis \citep{LATPaper}. 
The pile-up effect has been addressed in detail in \citet{2012ApJ...745..144A} and \citet{LATperform}. The list of all LAT detected flares, and the analysis of the 
first two flares with long lasting high-energy emission (2011 March 7--8 and 2011 June 7) is 
presented in \cite{0004-637X-787-1-15,0004-637X-789-1-20}. 

\subsection{June 2010: An impulsive event}

\begin{figure*}
\centering
\includegraphics[width=\linewidth]{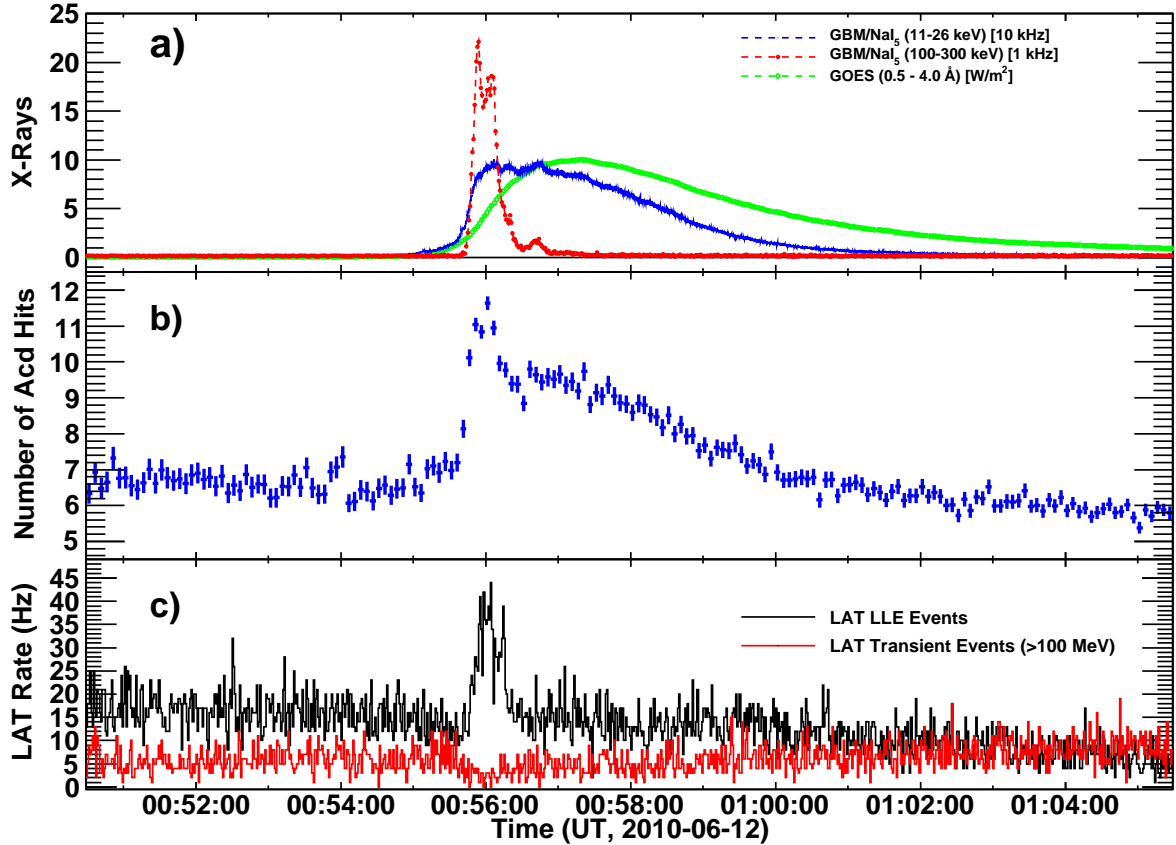}
\caption{
Time histories related to the 2010 June 12 solar flare.  a) {\it{GOES}} 0.5 -- 4 \AA rates, and GBM NaI 11 -- 26 keV and 100 -- 300 keV relative rates; b) LAT ACD hit rate $>$100 keV containing contributions from background, $>$100 keV solar flare X rays (impulsive peak) and pulse pile up from 10's of keV solar X rays following the NaI 11 -- 26 keV profile in 1a); and c) LLE and LAT Transient Class event rates. }
\label{lat_nai}
\end{figure*}

On 2010 June 12 00:30 UT a moderate {\it{GOES}} M2.0 class X-ray flare erupted from the active region (AR) 11081 located approximately N23$^{\circ}$W43$^{\circ}$. At the time of the flare the \Fermi spacecraft was in sunlight and during a relatively low-background portion of its orbit\footnote{The \Fermi observatory is in a nearly circular orbit with an inclination of 25.6$^{\circ}$ at 565 km.}. The GBM triggered on the flare at 00:55:05.64 UT and detected keV emission for $\approx$10 m. The $11 - 26$ keV emission recorded by the GBM NaI detectors rose precipitously for about 40 s and is shown in Figure \ref{lat_nai}a. For comparison we include the GOES 0.5 -- 4 \AA profile and note that this emission is dominated by 3 keV thermal photons as is reflected in its slower rise and extended tail. The $100 - 300$ keV time profile observed by the GBM's solar facing NaI detector is also plotted in Figure \ref{lat_nai}a. It is clear that the emission peaks more sharply and ends sooner at higher X-ray energies.

The accompanying hard X-ray emission from the flare was detected in the LAT's ACD and is reflected in the shape of the average number of ACD tile hits as a function of time (shown in Figure \ref{lat_nai}b). The broad peak with a maximum near 00:57 UT of the hit distribution has a shape similar to the 11 -- 26 keV emission and the impulsive peak is similar to the 100 to 300 keV flux observed by the GBM NaI detector. As shown by the red curve in Figure \ref{lat_nai}c there is no evidence for the flare in the well-screened standard LAT data products. \citep[What is shown here are the events belonging to the \texttt{P6TRANSIENT} event class,][]{atwo09}. This is the direct consequence of the pulse pile-up effect. The black curve in Figure \ref{lat_nai}c is the LAT LLE $>$30 MeV event rate for the time of the flare.

White light emission observed by the Helioseismic and Magnetic Imager (HMI) on the {\it{Solar Dynamics Observatory}} ({\it{SDO}}) \citep{oliv11} in a single 45 s exposure during the hard X-ray emission revealed two compact {\em footpoints} about 10$^4$ km apart.

The $>$30 MeV LLE spectrum of this flare revealed flare emission up to an 
energy of $\sim$400 MeV. The nuclear line emission observed with the GBM 
implies
the presence of accelerated ions up to at least 50 MeV nucleon$^{-1}$.  It is
possible that the flare-accelerated proton spectrum extended up to the
$\sim$300 MeV threshold for pion production.  Alternatively, it is also 
possible
that the LAT emission is from electron bremsstrahlung, either from an
extension to high energies of the electron spectrum producing the X-ray
bremsstrahlung observed in the GBM or from an additional hard electron
component.  One possible way to resolve this ambiguity is to jointly fit the
GBM and LAT spectra assuming different origins for the LAT emission.

\begin{figure}
\centering
\includegraphics[width=\linewidth]{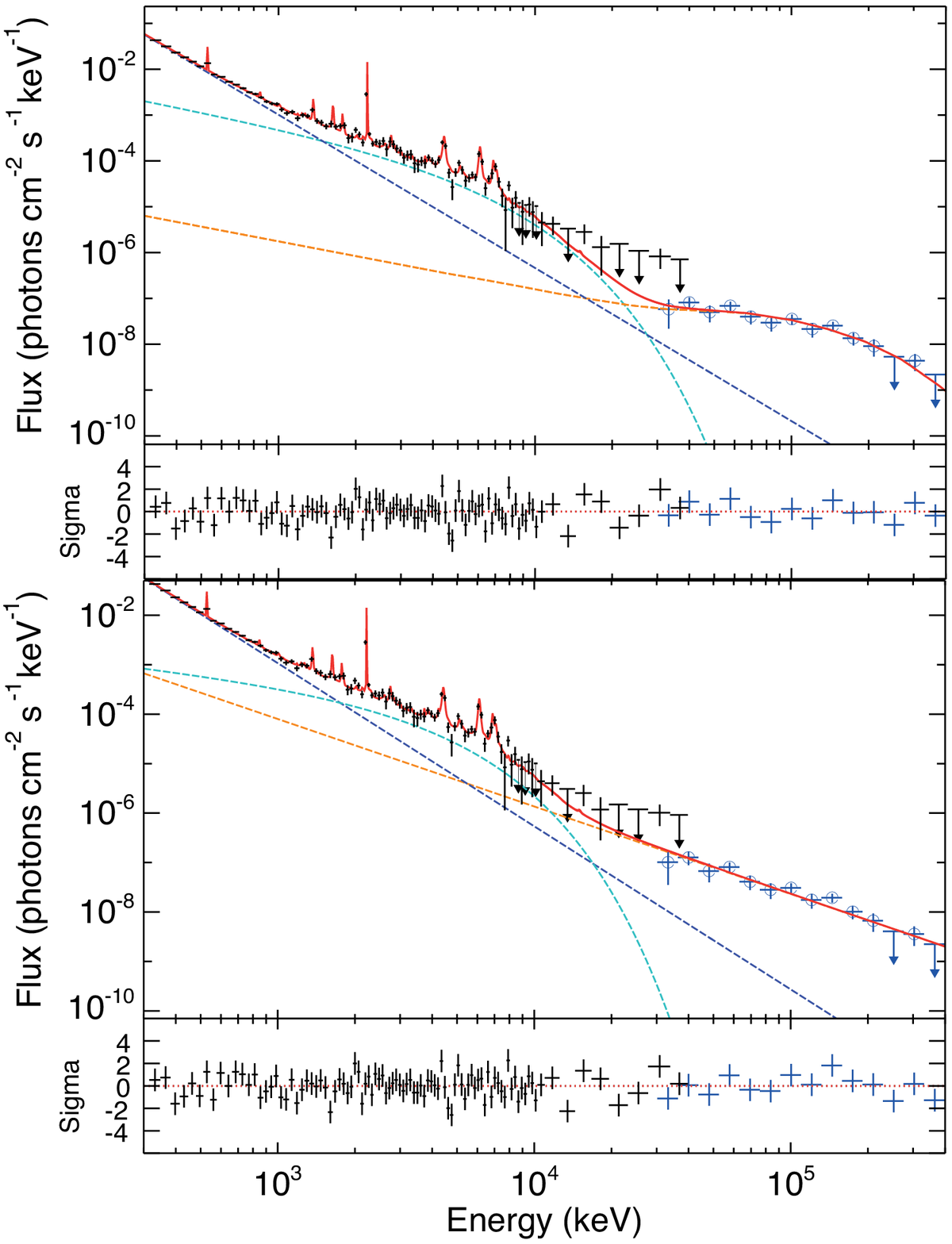}
\caption{
Combined GBM/LAT photon spectrum accumulated between 00:55:40 and 00:56:30
showing the best total fit using the same components as in Figure 3 plus an additional component for the LAT emission.  
The upper panel shows a pion-decay fit to the LAT spectrum; alternatively the lower panel shows a power-law
fit, presumedly representing a third electron bremsstrahlung component.
Note that because this is a photon representation the lines are plotted
at their intrinsic resolution and appear to be more significant than they really are.
}
\label{stack}
\end{figure}
In Figure \ref{stack} we plot the background-subtracted photon spectrum from
0.3 to 400 MeV including both the GBM and LAT data.  We made two fits, using {\em rmfit 3.4}\footnote[1]{R.S. Mallozzi, R.D. Preece, \& M.S. Briggs, ``RMFIT, A Lightcurve and Spectral Analysis Tool'', Robert D. Preece, University of Alabama in Huntsville, (2008): http://fermi.gsfc.nasa.gov/ssc/data/analysis/user/}, customized for the specific solar flare, and the {\em OSPEX}\footnote[2]{SolarSoft: http://www.lmsal.com/solarsoft/} analysis packages, to the joint data sets. In the first fit we assume that the observed LAT emission was from
pion-decay radiation (top panel of Figure \ref{stack}) and the other assuming that it was from a
hard power-law spectrum of electron bremsstrahlung (bottom panel).   Based on
the statistical quality of the fits to the LAT spectrum we cannot distinguish
between the two emission models but, if the LAT emission is from electron bremsstrahlung, we have found that it cannot be a simple extension of the low-energy bremsstrahlung components that we determined from fits to the GBM data; 
it must be from a distinct population of electrons extending to energies of several hundred MeV. 
However, this high energy electron component would produce a spectrum that steepens beyond tens of MeV due to synchrotron energy losses that increase with energy \citep[see][]{park97}, and must have a quite different origin.
Consequently we believe that this is a less likely scenario than the hadronic model.

%

Assuming that the LAT emission is from hadronic interactions, we have fit the LAT spectrum with calculated pion-decay templates~\citep{murp87}, which depends on the ambient density, composition and magnetic field, on the accelerated-particle composition, pitch angle distribution and energy spectrum. The templates represent a particle population with an isotropic pitch angle distribution and a power-law energy spectrum ($dN/dE \propto E^{-s}$, with $E$ the kinetic energy of the protons) interacting in a thick target with a coronal composition~\citep{ream95} taking  $^4$He/H = 0.1. 
With 67\% confidence (based on $\chi^2$) we conclude that the spectrum of accelerated ions responsible for the pion-decay emission must be steeper than a power-law with index $-$4.5.  We note that there is no change in the quality of the fits for indices steeper than $-5$ due to limited statistics $>$400 MeV. 
We can use the results of our GBM and LAT spectral analyses to obtain information on ions accelerated in the impulsive phase of the June 12 flare.  \citet{murp97} have described how parameters derived from integrated spectroscopic fits and temporal studies can be used to obtain this information.  We first use the nuclear de-excitation line, neutron-capture line, and pion-decay fluences to estimate the overall shape of the accelerated ion spectrum.  
These three emissions are produced by accelerated ions within distinct energy ranges: $\sim$5-20 MeV for the de-excitation lines, $\sim$10-50 MeV for the neutron capture line, and $>$300 MeV for the pion-decay emission.  
Ratios of these emissions therefore determine the relative numbers of accelerated ions in the associated energy ranges.  
We then obtain spectral indices across these energy ranges by comparing measured ratios with ratios from theoretical calculations \citep{murp87,murp05,murp07} based on updated nuclear cross sections.

If we assume that the LAT emission $>$30 MeV was entirely due to pion-decay emission, then we estimate that the flare-accelerated ion spectrum was consistent with a series of power laws, softening with energy, with indices of $\sim$$-3.2$ between $\sim5-50$ MeV, $\sim$$-4.3$ between $\sim$50--300 MeV, and softer than $\sim$$-4.5$ above 300 MeV. 
In Table~\ref{tab:tab1} we summarize our findings, reporting the processes responsible for the detected emission, energy range of emitted gamma-rays, as well as the energy and spectral index of the accelerated ions/electron distribution.

\begin{table}[t]
\begin{center}
\begin{tabular}{lccc}
\hline 
{Component} & {$\gamma$-rays} & {electrons/ions} & Spectral Index \\
& (MeV) & (MeV) & acc. particles\\
\hline 
\hline 
Brem.  & 0.1--1 & 0.1--1 & -3.2 \\
Brem.  & 2--10  & 2--10  & $<$-1.2\\
HE Brem.  & 10--200  & 10--200 & $\approx$-2.0\\
\hline
Neutron Capt.  & 2.2  & 5-50  & $\sim-$3.2\\
Nuclear lines  & 5-20  & 50-300  & $\sim-$4.3\\
Pions & $>$30  & \gtsima 300 & \ltsima$-$4.5\\
\hline
\hline
\end{tabular}
\end{center}
\caption{Derived quantities for accelerated particle distributions (with a cut-off at 2.4 MeV)}
\label{tab:tab1}
\end{table}

\subsection{March 2012: Impulsive and sustained emission of a bright flare}

\begin{figure*}
\includegraphics[width=\linewidth]{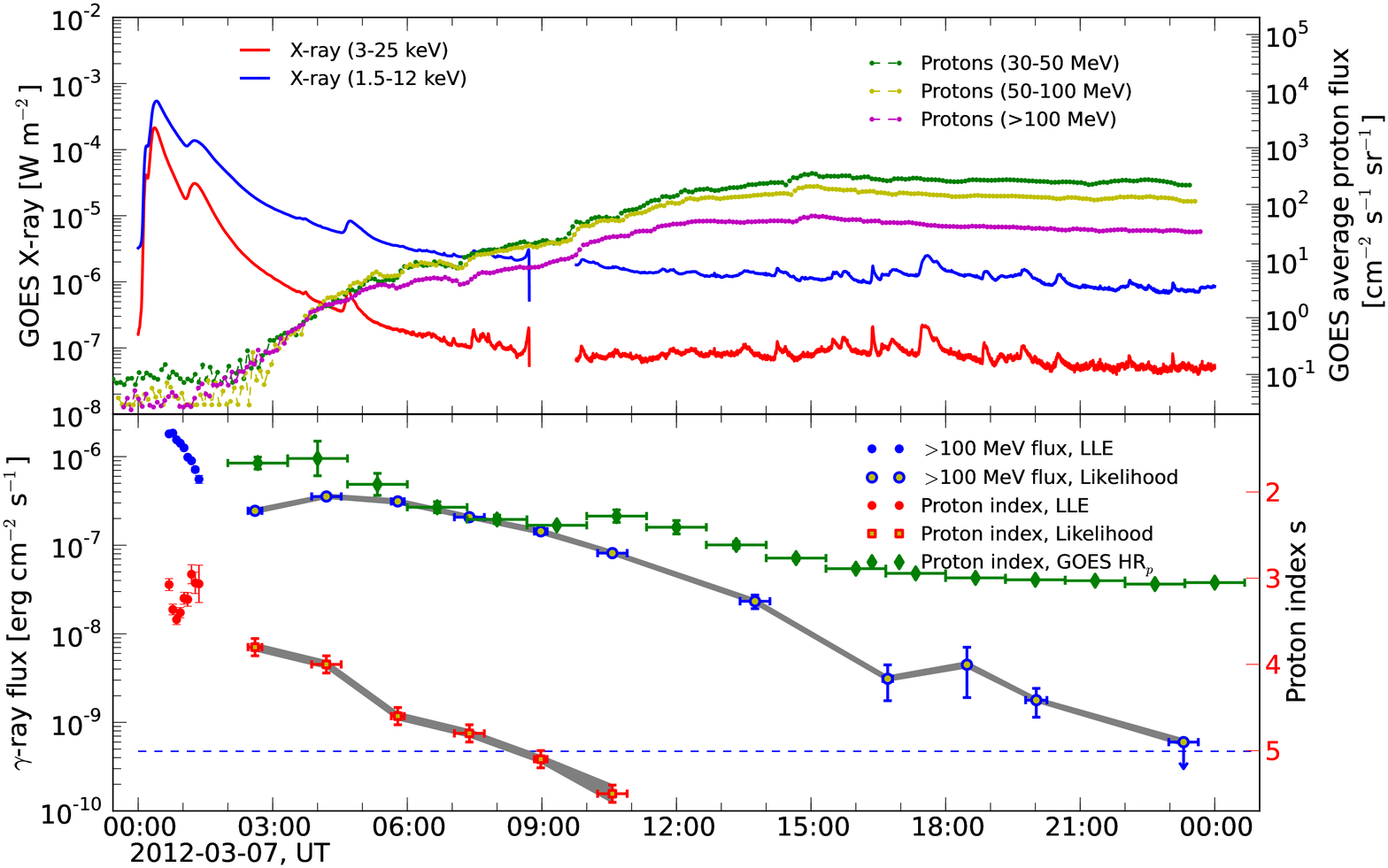}
\caption{Long lasting emission. {\bf Top panel}: soft X-rays (red: 1.5--12 keV, blue: 3--25 keV) from the {\it GOES} 15 satellite. On the right axis, 5-minute averaged proton flux (green: 30--50 MeV, yellow: 50--100 MeV, magenta: $>$100 MeV). We display the average of detectors A and B.
{\bf Bottom panel}: high-energy gamma-ray flux above 100 MeV measured by the \Fermi LAT. The Blue/red circles represent the flux and the derived proton spectral index obtained with the LLE analysis (covering the initial period, when the instrumental performance was affected by pileup of hard-X-rays in the ACD tiles). 
The blue circles and red squares represent the flux and derived proton spectral index, respectively, obtained by standard likelihood analysis. Green diamonds are the {\it GOES} proton spectral indexes derived from the hardness ratio, as described in the text.
The gray bands correspond to the systematic uncertainty associated with flux measurements and of the estimated proton index due to uncertainties on the effective area of the instrument. The horizontal dashed line corresponds to the value of the gamma-ray flux from the quiescent Sun, from \citet{2011ApJ...734..116A}.}
\label{March2012}
\end{figure*}

On 2012 March 7 two bright X-class flares originating from the AR NOAA AR\#:11429 (located at N16$^{\circ}$E30$^{\circ}$) erupted within an hour of each other, marking one of the most active days of Solar Cycle 24. 
The first flare started at 00:02:00 UT and reached its maximum intensity (X5.4) at 00:24:00 UT while the second X1.3 class flare occurred at 01:05:00 UT, reaching its maximum 9 minutes later. 

The GOES satellite observed intense X-ray emission beginning at about 00:05:00 UT and lasting for several hours. 
Moreover, it detected Solar Energetic Particles (SEP) protons in three energy bands originating these flares. 
The Reuven Ramaty High-Energy Solar Spectroscopic Imager \citep[RHESSI,][]{2002SoPh..210....3L} was not observing the Sun during this period.
On the top panel of Figure~\ref{March2012} we plot the X-ray data from GOES 15 satellite measured in both 3--25 keV and 1.5--12 keV, as well as the detected proton flux.


The \Fermi LAT $>$100 MeV count rate was dominated by the gamma-ray emission from the Sun\footnote{http://apod.nasa.gov/apod/ap120315.html}, which was nearly 100 times brighter than the Vela Pulsar in the same energy range.
During the impulsive phase (the first eighty minutes) the X5.4 flare was so intense that the LAT observation suffered from the pile-up effect so we used the LLE technique to analyze the impulsive phase of this bright flare.

We fit the data using \texttt{XSPEC}\footnote{http://heasarc.gsfc.nasa.gov/docs/xanadu/xspec/index.html} to test three models. The first two are simple phenomenological functions, to describe bremsstrahlung emission from accelerated electrons, namely a pure power law (PL) and a power law with an exponential cut-off (EXP):
\begin{equation}
 \frac{dN(E)}{dE} = N_{0}\,\e^{-\Gamma}\,\exp\left({-\frac{E}{E_{co}}}\right);
\label{eq1}
\end{equation} 
where $\Gamma$ is the photon index and $E_{co}$ is the cut-off energy.
We found that the data clearly diverge from a pure power law spectrum and that the EXP provides a better fit in all time intervals considered.
The third model used the same pion decay templates ~\citep{murp87} used for the 2010 June 2 flare.
When using the pion-decay templates to obtain the gamma-ray flux value we fit the data varying the proton spectral index from 2 to 6, in steps of 0.1. 
In this way, we fit the LAT data with a model with two free parameters, the normalization and the proton index $s$.


To study the temporally-extended emission, we perform time-resolved spectral analysis in Sun-centered coordinates by transforming the reference system from celestial 
coordinates to ecliptic Sun-centered coordinates. 
This is necessary in order to compensate for the effect of the apparent motion of the Sun during the long  duration of the flare. 
We select intervals when the Sun was in the FOV (angular distance from the LAT boresight $<$ 70$^{\circ}$)
and use the unbinned maximum likelihood algorithm implemented in \texttt{gtlike}\footnote{We used \texttt{ScienceTools} version 09-28-00 available from the \Fermi Science Support Center \url{http://fermi.gsfc.nasa.gov/ssc/}}. 

We include the isotropic template model that is used to describe the extragalactic gamma-ray emission and the 
residual cosmic ray (CR) contamination\footnote{We used \texttt{iso\_p7v6source.txt} available from the \Fermi Science Support Center}, leaving its normalization as the free parameter.
Over short time scales, the diffuse Galactic emissions produced by CR interacting with the interstellar medium are not spatially resolved and are hence included in the isotropic template.
We also add the gamma-ray emission from the quiescent Sun modeled as a point source located at the center of the disk, with a spectrum described by a simple power law with a spectral index of 2.11 and an integrated energy flux ($>$ 100 MeV) of 4.7$\times$10$^{-10}$ erg\,cm$^{-2}$\,s$^{-1}$ \citep[corresponding to a flux of 4.6$\times$10$^{-7}$ ph\,cm$^{-2}$\,s$^{-1}$ as reported in][]{2011ApJ...734..116A}.
We did not include the extended Inverse Compton (IC) component described in \citet{2011ApJ...734..116A} because it is too faint to be detected during these time intervals.
We fit the data with the same two phenomenological functions used for the impulsive phase of the flare and use the likelihood ratio test to estimate whether the addition of the exponential cut-off is statistically significant.
The Test Statistic (TS)~\citet{Mattox:96} is defined as twice the increment of the logarithm of the likelihood $\like$ obtained by fitting the data with the source and background model components simultaneously.
Because the null hypothesis (i.e. the model without an additional source) is the same for the two models, the increment of the TS ($\Delta$TS=TS$_{\rm PLEXP}$-TS$_{\rm PL}$) is equivalent to the corresponding difference of maximum likelihoods computed between the two models. 

For each interval, if $\Delta$TS $\geq$ 30 (roughly corresponding to 5$\sigma$) then the PLEXP model provides a significantly better fit than the simple power-law and we retain the additional spectral component. 
In these time intervals, we also used the pion decay model to fit the data and estimated the corresponding proton spectral index.
We performed a series of fits with the pion decay template models calculated for a range of proton spectral indices. 
We then fit the resulting profile of the log-likelihood function with a parabola and determine its minimum ($\like_{\rm min}$) and the corresponding value $s_0$ as the maximum likelihood value of the proton index. 

In the lower panel of Figure~\ref{March2012} we combine the LLE and likelihood analysis results, showing the evolution of  both the  gamma-ray flux and the derived spectral index of the protons\footnote{After approximately 11:00:00 UTC the flux of the Sun diminished to the point that the spectral index of the proton distribution cannot be significantly constrained.}.
In the last five time intervals the power-law representation is adequate to describe the data; in the last bin, the flare is only marginally significant (TS=7); the flux and the photon index are compatible with the values of the quiescent Sun. For this reason we have indicated the last point as an upper limit (computing the 95\% C.L.). Unlike during the impulsive phase, the spectrum during the temporally extended phase becomes softer monotonically ($s$ increases).

We also compare our results with the {\it GOES} proton spectral data. For this, we selected two energy bands ($>$30 MeV and $>$100 MeV) and corrected the light curve by the proton time-of-flight (TOF) to 1 AU by considering the TOF for 30 MeV and 100 MeV protons (i.e. the maximum delay in each energy band).
As a measure of the spectral index of the SEP protons ($s_{\rm SEP}$), we compute the Hardness Ratio HR$_{p}$ defined as:
\begin{equation}
{\rm HR}_{p}=\ln\frac{\rm P_{>100 MeV}}{\rm P_{>30MeV}}
\end{equation}
where P is the integral of the proton flux (assuming that the proton flux is proportional to a power law). The HR$_{p}$ is related to the value of the spectral index, $s_{\rm SEP}$, of the SEP protons observed at 1 AU, roughly as:
\begin{equation}
s_{\rm SEP} \sim 1-0.83\,{\rm HR}_{p}
\end{equation}

To estimate the uncertainty associated with this procedure we repeat the calculation neglecting the TOF correction. In this way we obtain two values for the SEP spectral index for each time bin, corresponding to the actual and zero delay due to the time of flight. In Figure~\ref{March2012} we report the estimated proton spectral index as the average of these two values and its uncertainty as half the difference of these two values. However we note that the $s_{\rm SEP}$ is for protons with energy less than a few hundred MeV while $s$ is for protons with energies greater than 300 MeV. Diffusion is expected to play an important role in the transport of these SEPs but an in-depth transport analysis is beyond the scope of this paper. From our comparison we find that the proton spectral index inferred from the gamma-ray data is systematically softer than the value of the index derived directly from SEP observations but that the temporal evolution (hard-to-soft) is similar.

Uncertainties in the calibration of the LAT introduce systematic errors on the measurements. The uncertainty of the effective area is dominant, and for the \texttt{P7SOURCE\_V6} event class it is estimated to be $\sim$10\% at 100 MeV, decreasing to $\sim$5\% at 560 MeV, and increasing to $\sim$10\% at 10 GeV and above. We studied the effect of the systematic uncertainties on our final results via the bracketing technique described in detail in \citet{2012ApJS..203....4A}. We find that the uncertainties on the flux are $<$10\% and on the inferred proton index are $<$0.10. The results are represented by the gray bands in Figure~\ref{March2012}.

\section{New observations: the behind-the-limb synopsis}

\begin{figure}[htb!]
\begin{center}
\includegraphics[width=1.0\linewidth]{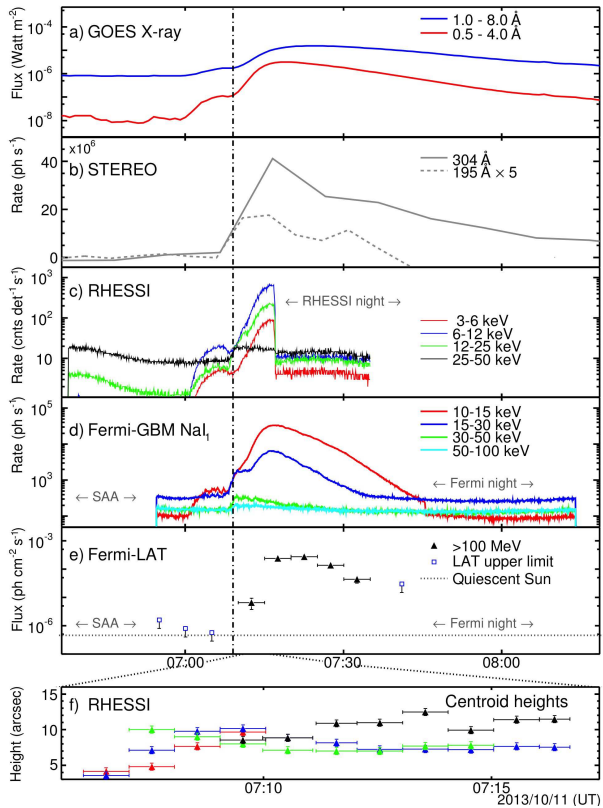}\\
\caption{{\bf PRELIMINARY:} Light curves of the 2013 October 11 flare as detected by a) \emph{GOES}, b) \stereo, c) \hsi, d) GBM, e) LAT and e) \hsi emission centroid heights, with the same color coding as in c). \Fermi exited the South Atlantic Anomaly (SAA) at 06:57:00 UT.} 
\label{fig:LightCurve}
\end{center}
\end{figure}

On 2013 October 11 at 07:01 UT a GOES M1.5 class flare occurred with soft X-ray emission lasting 44 min and peaking at 07:25:00 UT. Figure~\ref{fig:LightCurve} shows the GOES, \stereo-B, \hsi, \Fermi GBM and LAT lightcurves of this flare. LAT detected $>$100~MeV emission for $\sim$30 min with a peak flux between 07:20:00--07:25:00 UT. \hsi coverage was from 07:08:00 $-$ 07:16:40 UT, overlapping with \Fermi for 9 min.

\begin{figure}[ht]
\begin{center}
\includegraphics[width=1.0\linewidth]{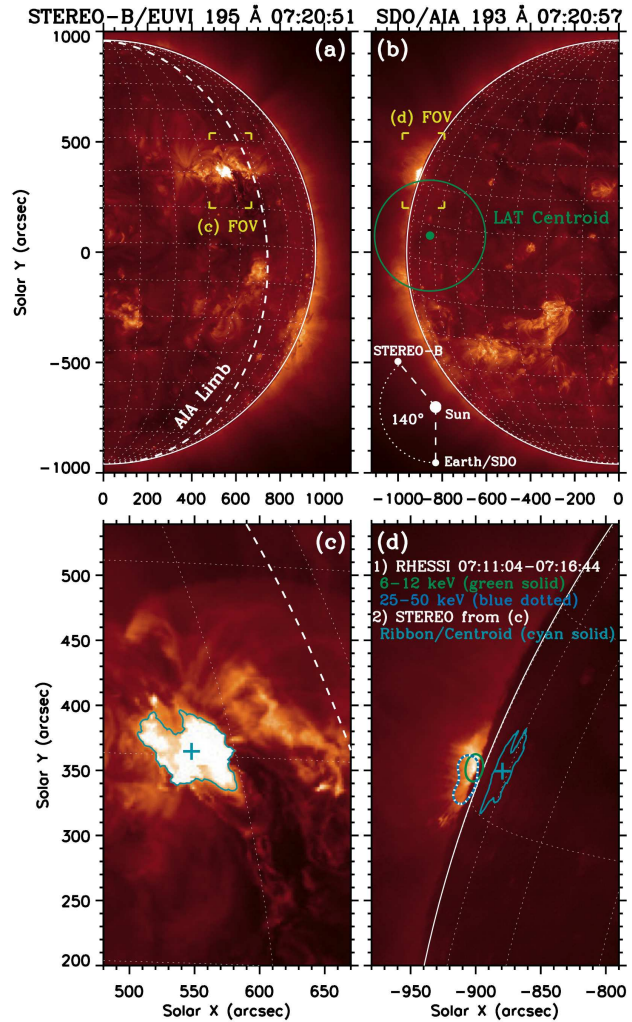}\\
\caption{{\bf PRELIMINARY:} \stereo-B (left) and \sdo (right) images near the flare peak. The white-dashed line in (a) and (c) represents the solar limb as seen by \sdo. The green line in (b) shows the 68\% error circle for the LAT source centroid. The cyan contour and plus sign in (c) mark the {\it STEREO} flare ribbon and its centroid, respectively. Their projected view as seen from the AIA perspective is shown in (d), in which the centroid is located at 9.9\de behind-the-limb. The green and blue-dotted contours in (d) show \hsi sources. The rectangular brackets in (a) and (b) mark the field of view (FOV) for (c) and (d), respectively.}
\label{fig:STEREOSDO}
\end{center}
\end{figure}

Images in Figure~\ref{fig:STEREOSDO} from the \stereo-B Extreme-UltraViolet Imager~\citep[EUVI;][]{EUVIinstrument} and the \sdo Atmospheric Imaging Assembly~\citep[AIA;][]{AIApaper} of the photosphere indicate that the  AR was $\sim$9.9\de behind the limb at the time of the flare. LASCO onboard the \emph{Solar and Heliospheric Observatory} (\soho) observed a backside asymmetric halo CME associated with this flare beginning at 07:24:10 UT with a linear speed of 1200 km/s~\citep{CMEcatalog} and a bright front over the Northeast. Both \stereo\ spacecrafts detected energetic electrons, protons, and heavier ions including helium, as well as type-II radio bursts indicating the presence of a coronal--heliospheric shock. \sdo observed a global EUV wave (Liu et al. 2015, in prep.), possibly the coronal counterpart of the shock. \stereo-B had an unblocked view of the entire flare and detected a maximum rate of 3.5$\times$10$^{6}$ photons/sec in its 195 \Angst\ channel, corresponding to a GOES M4.9 class~\citep{2013SoPh..288..241N} if it had not been occulted.

The LAT data were analyzed using the unbinned maximum likelihood algorithm \texttt{gtlike} implemented in the \Fermi \texttt{ScienceTools}\footnote{We used version 09-30-01 available from the \Fermi Science Support Center \url{http://fermi.gsfc.nasa.gov/ssc/} } with \texttt{P7REP\_SOURCE\_V15} instrument response functions. We selected gamma-rays from a 12\de region centered on the Sun and within 100\de of the zenith to reduce contamination from the Earth's limb.  For \hsi data, we implemented the CLEAN imaging algorithm~\citep{Hurford2002} using the detectors 3$-$9 to reconstruct the X-ray images.  We used the FITS World Coordinate System software package~\citep{2010SoPh..261..215T} to co-register the flare location between \stereo\ and \sdo images. The \stereo\ light curves are pre-flare background subtracted, full-Sun integrated photon rates.

We measure the direction of the LAT $>$ 100 MeV gamma-ray emission \citep[as described in][]{0004-637X-789-1-20} and find a best fit position for the emission centroid at heliocentric coordinates of ($-855''$,$75''$) with a 68\% error radius of 251$''$, as shown in Figure~\ref{fig:STEREOSDO}(b). \hsi\ X-ray sources integrated over 07:11:04$-$07:16:44~UT are shown as 80\%-level, off-limb contours in Figure~\ref{fig:STEREOSDO}(d). 

The temporal evolution of the projected \hsi source heights above the solar limb are shown in Figure~\ref{fig:LightCurve}(f). The higher-energy emission generally comes from greater heights, consistent with expectations for a loop-top source~\citep[e.g.,][]{Masuda1994Nature, SuiL2003ApJ...596L.251S, 2004ApJ...611L..53L}. If this were a footpoint source, we would expect an opposite trend since larger column depths are required for stopping higher-energy electrons \citep[e.g.,][]{LiuW2006ApJ...649.1124L, kontaretal}. Moreover, from \sdo/AIA movies we find no signature of EUV ribbons, even in the late phase during the \hsi night. Together, these observations provide convincing evidence that the footpoints were indeed occulted. 

\section{Discussion}
We have presented the analysis of three solar flares detected by the \fermi LAT at high energy, and we highlight some of the similarities and differences of these flares. 
The high-energy emission of the 2012 June solar flare seems to be correlated with HXR emission, suggesting that acceleration of particles and gamma-ray emission take place close in space. Specifically, particles accelerated at the loop top could propagate along the loop field lines interacting and emitting gamma-rays at the footprint. For this flare, there is no evidence for precipitation of trapped flare particles, particles accelerated in magnetic loops after the impulsive phase, particles accelerated in CME-associated reconnecting current sheets \citep{ryan00}, or particles sharing the same origin as the Solar Energetic Particles (SEPs) observed in space \citep{ram87,cliv93}. 
On the other hand, flares with long (or sustained) gamma-ray emission have also been observed by the \fermi LAT. Temporal and spectral analysis suggests that, even if the short impulsive phase is clearly visible at $>$ 100 MeV energies, the sustained long lasting emission is more correlated with SEP properties, suggesting that, for this class of flares,  either long trapping, continuous acceleration, or acceleration at the CME shock could be a better explanation. 
The behind-the-limb flare detection at high-energy adds additional considerations that are extremely useful for understanding the physics of particle acceleration and gamma-ray production during solar flares.
We have presented preliminary results from the 2013 October 11 solar flare from \fermi, \hsi, \sdo and \stereo. \stereo-B images indicate that the flare occurred in an AR 9.9\de behind-the-limb. \hsi and GBM NaI$_{1}$ detected HXRs up to 50 keV from the flaring loop-top. The most unusual aspect of this flare is the LAT detection of photons of energies $\epsilon >$100 MeV for about 30 minutes with some photons having energies up to several GeV. 
\begin{figure}[ht]
\begin{center}
\includegraphics[width=0.4\textwidth, trim=0cm 0cm 0cm 0cm]{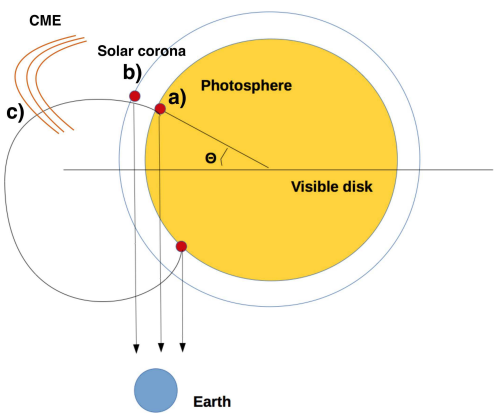}
\caption{Model a): acceleration at the flare, gamma-ray emission site below the photosphere; b) acceleration at the flare, gamma-ray emission in the corona above the limb; c) acceleration (or re-acceleration) at the CME-shock, gamma-ray emission at the Sun.}
\label{fig:models}
\end{center}
\end{figure}

We consider three scenarios for the emission site of the gamma-rays, outlined in Figure \ref{fig:models}. 
Electrons or protons with energies $E > \epsilon$ can produce these photons after traversing a column depth of matter which is much larger than the depth penetrated by HXR producing electrons (model a). For occulted flares the emitted photons must traverse even larger depths than particles and they may be scattered and absorbed. 
Alternatively, acceleration and gamma-ray emission can take place in the corona above the limb (model b), suggesting trapping of particles, e.g., by strongly converging magnetic fields. In the third model (model c) CME-shock accelerated particles can travel back to the Sun along magnetic fields connecting the acceleration site with the visible side of the Sun. \fermi LAT observation of the  2013 October 11 flare (paper in preparation) shows that model a) can be ruled out.
The LAT detection of gamma-ray emission from a flare with $\theta>20$\de also poses some complications to the second scenario (model b), as particles will have to be accelerated even further away in the corona, where densities are very low. Acceleration (or re-acceleration) at the CME shock (model c) remains possible. 
\fermi LAT observations are becoming very important to disentangle models of particles acceleration and gamma-ray production in solar flares. 
Future LAT observations, combined with a systematic study of the solar flares detected at high energy, will very likely help to understand this fascinating problem, as well as to improve our knowledge of particle acceleration in astrophysical sources in general. 

\begin{acknowledgments}
The $Fermi$ LAT Collaboration acknowledges support from a number of agencies and institutes for both development and the operation of the LAT as well as scientific data analysis. These include NASA and DOE in the United States, CEA/Irfu and IN2P3/CNRS in France, ASI and INFN in Italy, MEXT, KEK, and JAXA in Japan, and the K.~A.~Wallenberg Foundation, the Swedish Research Council and the National Space Board in Sweden. Additional support from INAF in Italy and CNES in France for science analysis during the operations phase is also gratefully acknowledged.
\end{acknowledgments}

\bigskip
\bibliography{Omodei_SolarFlares}
\end{document}